\documentclass[conference]{IEEEtran}

\usepackage{algorithm}
\usepackage{algpseudocode}
\usepackage{amssymb}
\usepackage{amstext}
\usepackage{amsmath}
\usepackage{amsthm}
\usepackage{multicol}
\usepackage{pgfplots}
\usepackage{pslatex}
\def\BibTeX{{\rm B\kern-.05em{\sc i\kern-.025em b}\kern-.08em
		T\kern-.1667em\lower.7ex\hbox{E}\kern-.125emX}}

\begin{document}

\title{De(con)struction of the lazy-F loop: improving performance of Smith Waterman alignment}

\author{\IEEEauthorblockN{Roman Snytsar}
	\IEEEauthorblockA{\textit{Research and AI} \\
		\textit{Microsoft}\\
		Redmond, USA \\
		Roman.Snytsar@microsoft.com}}
	
\maketitle

\begin{abstract}
Striped variation of the Smith-Waterman algorithm is known as extremely efficient and easily adaptable for the SIMD architectures. However, the potential for improvement has not been exhausted yet. The popular Lazy-F loop heuristic requires additional memory access operations, and the worst-case performance of the loop could be as bad as the nonvectorized version. We demonstrate the progression of the lazy-F loop transformations that improve the loop performance, and ultimately eliminate the loop completely. Our algorithm achieves the best asymptotic performance of all scan-based SW algorithms O(n/p+log(p)), and is very efficient in practice.	
\end{abstract}

\begin{IEEEkeywords}
	Parallel Processing, Vectorization, Bioinformatics, Smith-Waterman
\end{IEEEkeywords}

\section{Background}

Sequence alignment is an essential component of many bioinformatics data processing workflows. Smith-Waterman algorithm (SW) produces an optimal local alignment between the two sequences. At the same time it is one of the slowest sequence alignment algorithms used, due to its quadratic computational complexity. Aligning two sequences of the lengths $m$ and $n$ requires $O(mn)$ time. This makes the SW algorithm a continual focal point of optimization attempts using all kinds of the hardware accelerators.

Farrar \cite{b5} has proposed the \emph{Striped} algorithm that uses the SSE vector instructions over an array of vectors containing evenly spaced data cells along the length of the input sequence. Striped approach eliminates most of the dependencies between the vectors in the array, and ignores the remaining dependencies during the first pass through the array. An additional pass, known as the Lazy-F update loop, is required to account for the inter-vector dependencies. However, it is often possible to break from the lazy-F loop early, making the Striped algorithm an excellent performer.

Algorithm \ref{al:sw} shows the striped algorithm with the lazy-F loop adaptation from \cite{b9}. Here operator $\lll$ denotes the vector shift, and operator $-$ indicates subtraction with saturation as none of the vector values can be negative. All vector variable names are PascalCased while all scalar variable names are camelCased.
\begin{algorithm}
	\caption{Striped Smith-Waterman}\label{al:sw}
	\begin{algorithmic}[0]
		\Procedure{SmithWaterman}{}
		\State$segLen\gets (length(query) + p-1)/p$
		\For{$i = 0\dots refLen$}
		\State$F \gets 0$  
		\State$H \gets HStore[segLen - 1] \lll 1$
		\State$swap (HLoad, HStore)$
		
		\For{$j = 0\dots segLen$}
		\State$H \gets H + vprofile[i][j]$
		\State$Max \gets max(Max, H)$
		\State$H \gets max (H, E[j])$ 
		\State$H \gets max (H, F)$
		
		\State$HStore[j] \gets H$
		
		\State$H \gets H - GapOpen$
		\State$E[j] \gets E[j] - GapExtend $
		\State$E[j] \gets max (E[j], H)$
		
		\State$F \gets F - GapExtend$
		\State$F \gets max (F, H)$
		
		\State$H \gets HLoad[j]$ 
		\EndFor

		\For{$k = 1\dots p$} //Lazy-F Loop
		\State$F \gets F \lll 1$   
		\For{$j = 0\dots segLen$}
		\State$HStore[j] \gets max (HStore[j], F)$
		\State$F \gets F - GapExtend$
		\If{$(F-H)==0$}
		\State\textbf{break}
		\EndIf
		\EndFor
		\EndFor
		\EndFor 
		\EndProcedure
	\end{algorithmic}
\end{algorithm}

With the input sequence length $n$ and vector size $p$ the computational complexity of the both passes comprising the outer loop iteration is $O((1+C)*n/p)$ where $C$ is a corrective factor describing the "laziness" of the second pass. The detailed analysis in \cite{b3} has shown that $c$ could vary significantly depending on the vector width, query length, and the query data. We will demonstrate the real world example where simply increasing the vector width changes $C$ by the factor of 10.

An alternative approach to linearizing the data dependencies between the vectors has been presented in \cite{b7} for their GPU SW implementation. The correction pass is formulated in terms of the parallel scan operation requiring additional $log(n)$ steps \cite{b1}. The overall computational complexity of the outer loop iteration is guaranteed to be $O(n/p+log(n))$.

In \cite{b4} the parallel scan has been translated to SIMD instructions. However, the computational complexity of that implementation is limited to $O(n/p+p)$.

The performance comparisons of the two SIMD approaches have revealed no clear winner \cite{b3}. In tests of different query lengths and vector widths the advantage shifts from striped to scan and back. This is partially due to the fact that the lazy-F is a heuristic and, while fast for many real world inputs, gives no performance guarantees.

\section{\uppercase{Solution}}
We will develop a hybrid approach that uses the striped layout but replaces the correction
 loop with the scan pass. Our work was inspired by \cite{b7}, but instead of the multi-page arithmetic proof, we will arrive at scan by performing a series of the equivalent program transformations of the lazy-F loop.
 
We start with removing the early loop exit condition for clarity (Algorithm \ref{al:lazyf}). 

\begin{algorithm}
	\caption{Lazy-F loop}\label{al:lazyf}
	\begin{algorithmic}[0]
		\For{$k = 1\dots p$}
		\State$F \gets F \lll 1$   
		\For{$j = 0\dots segLen$}
		\State$HStore[j] \gets max (HStore[j], F)$
		\State$F \gets F - GapExtend$
		\EndFor
		\EndFor
	\end{algorithmic}
\end{algorithm}

Note that in the lazy-F loop there is no dependency between the values of HStore[j], and, using the associative property of operator $max$, we can separate the F and H loops by storing the corrected F  vector for every segment (Algorithm \ref{al:fhsep}). 
\begin{algorithm}
	\caption{F and H loop separation}\label{al:fhsep}
	\begin{algorithmic}[0]
		\For{$k = 1\dots p$}
		\State$F \gets F \lll 1$   
		\For{$j = 0\dots segLen$}
		\State$FStore[j] \gets max (FStore[j], F)$
		\State$F \gets F - GapExtend$
		\EndFor
		\EndFor
		\For{$j = 0\dots segLen$}
		\State$HStore[j] \gets max (HStore[j], F)$
		\EndFor		
	\end{algorithmic}
\end{algorithm}

Operators $\lll$ and $-$ are mutually associative in a sense that $(A \lll n) - cI = (A-cI)\lll n$ where $I$ is a unity vector. Using that we can swap the inner and outer loops for F (Algorithm \ref{al:loopinv}). 

\begin{algorithm}
	\caption{Inner/outer loop inversion}\label{al:loopinv}
	\begin{algorithmic}[0]
		\For{$j = 0\dots segLen$}
		\State$Fj \gets F$   		
		\For{$k = 1\dots p$}
		\State$Fj \gets Fj \lll 1$   
		\State$FStore[j] \gets max (FStore[j], Fj)$
		\State$Fj \gets Fj - segLen * GapExtend$
		\EndFor
		\State$F \gets F - GapExtend$
		\EndFor
		\For{$j = 0\dots segLen$}
		\State$HStore[j] \gets max (HStore[j], F)$
		\EndFor		
	\end{algorithmic}
\end{algorithm}

Now it is obvious that the input vectors for every inner F loop invocation differ only by a constant vector, so the results can be calculated simply by subtracting the same constant from the result of the first and only inner loop execution. Note the change in the H loop of Algorithm \ref{al:loopelim}.

\begin{algorithm}
	\caption{Outer loop elimination}\label{al:loopelim}
	\begin{algorithmic}[0]
		\State$Fj \gets 0$   		
		\For{$k = 1\dots p$}
		\State$F \gets F \lll 1$   
		\State$Fj \gets max (Fj, F)$
		\State$F \gets F - segLen * GapExtend$
		\EndFor
		\For{$j = 0\dots segLen$}
		\State$HStore[j] \gets max (HStore[j], F)$
		\State$Fj \gets Fj - GapExtend$		
		\EndFor		
	\end{algorithmic}
\end{algorithm}

The loop for computing F can be replaced by the parallel scan operation, further reducing the number of operations from $p$ to $log(p)$. Actually, this scan is very similar to one in \cite{b7} indicating that our algorithm exploits the same optimization but for the striped version of the algorithm. The remaining H update loop can be executed even lazier, in the next iteration of the main loop, thus eliminating the lazy-F loop completely save for the scan with the $log(p)$ execution time (Algorithm \ref{al:scan}). 

\begin{algorithm}
	\caption{Striped Smith-Waterman Scan}\label{al:scan}
	\begin{algorithmic}[0]
		\Procedure{SmithWatermanScan}{}
		\State$segLen\gets (length(query) + p-1)/p$
		\For{$i = 0\dots refLen$}
		\State$F \gets 0$  
		\State$Fj \gets 0$   		
		\State$H \gets HStore[segLen - 1]$
 		\State$H \gets max (H, Fj - (segLen-1) * GapExtend)$
		\State$H \gets H \lll 1$
		\State$swap (HLoad, HStore)$
		
		\For{$j = 0\dots segLen$}
		\State$H \gets H + vprofile[i][j]$
		\State$Max \gets max(Max, H)$
		\State$H \gets max (H, E[j])$ 
		\State$H \gets max (H, F)$
		
		\State$HStore[j] \gets H$
		
		\State$H \gets H - GapOpen$
		\State$E[j] \gets E[j] - GapExtend $
		\State$E[j] \gets max (E[j], H)$
		
		\State$F \gets F - GapExtend$
		\State$F \gets max (F, H)$
		
		\State$H \gets HLoad[j]$ 
		\State$H \gets max (H, Fj)$ 
		\State$Fj \gets Fj - GapExtend$		
		\EndFor

		\State\textbf{for} $k = 1\dots p$ \textbf{scan}
		\State\quad$F \gets F \lll 1$   
		\State\quad$Fj \gets max (Fj, F)$
		\State\quad$F \gets F - segLen * GapExtend$
		\State\textbf{end scan}
		\EndFor 
		\EndProcedure
	\end{algorithmic}
\end{algorithm}
The asymptotic complexity of the outer loop iteration is $O(n/p+log(p))$, making our scan-based algorithm the fastest asymptotically.
\section{\uppercase{Implementation}}
We have implemented the scan algorithm using the SSE4.1, AVX2, and AVX512 instruction set. Later we have added another implementation that uses all 32 vector registers of the AVX512 architecture but issues only 256-bit wide instructions from the AVX2 and AVX512VL instruction sets. We will discuss its advantages in the results section below.
\subsection{Experimental Setup}
The computer platform is an Intel Xeon Platinum 8168 system with 16 cores running at 2.7 GHz and 32GB of RAM.
To test the software performance we have run the BWA alignment tool\cite{b8} with 16 threads (-t 16) on a 30X Human genome sample NA12878 from the 1000 Genomes database\cite{b2} using hg38 as a reference. We have executed the BWA version 7.15 to establish the baseline, and then replaced the Smith-Waterman mate rescue code with our SSE, AVX2 and AVX512 implementations. Additionally, we have extended the Lazy-F implementation to the AVX2 and AVX512 architectures. The total run times got collected from the BWA reports, and the percentage of time spent in the SW routine was measured via profiling. The alignment for every configuration has been run 3 times, and presented numbers are the averages of the observed run times.

\subsection{Results}

\begin{figure}[!tpb]
	\begin{tikzpicture}
	\begin{axis}[
	ybar=8pt,
	symbolic x coords={SSE,AVX2,AVX512,AVX512VL},
	xtick=data,
	nodes near coords,
	nodes near coords align={vertical},
	ylabel=Time \, mins]
	\addplot coordinates {
	(SSE,2.76)
	(AVX2,1.90)
	(AVX512,1.91)
	(AVX512VL,1.63)
    };
	\addplot coordinates {
		(SSE,2.72)
		(AVX2,2.03)
		(AVX512,17.9)
	};
	\legend{Scan, Lazy-F}
	\end{axis}
	\end{tikzpicture}
	\caption{Performance of the SW implementations.}\label{swperf}
\end{figure}
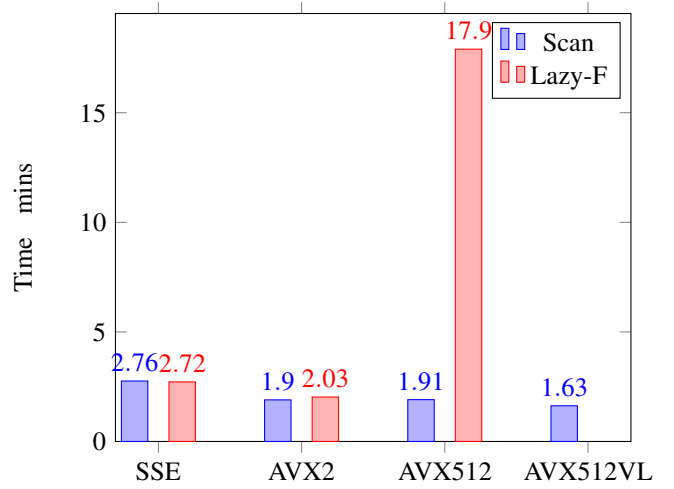
Figure \ref{swperf} shows the times spent in the SW routine for the Scan and Lazy-F algorithm implementations on the various vector width architectures. The obvious standout is the poor performance of the 
AVX512 lazy-F implementation. This early observation commensurates with the results in \cite{b3}, and it has motivated us to search for better solution in the first place. 

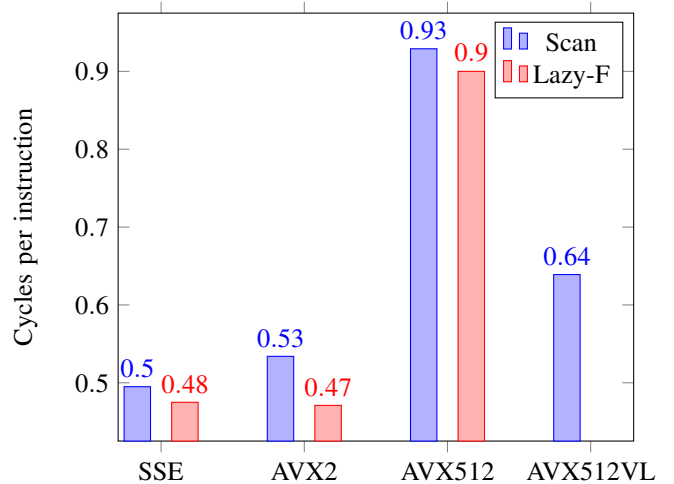
\begin{figure}[!tpb]
	\begin{tikzpicture}
	\begin{axis}[
	ybar=8pt,
	symbolic x coords={SSE,AVX2,AVX512,AVX512VL},
	xtick=data,
	nodes near coords,
	nodes near coords align={vertical},
	ylabel=Cycles per instruction]
	\addplot coordinates {
		(SSE,0.495)
		(AVX2,0.534)
		(AVX512,0.929)
		(AVX512VL,0.639)
	};
	\addplot coordinates {
		(SSE,0.475)
		(AVX2,0.471)
		(AVX512,0.9)
	};
	\legend{Scan, Lazy-F}
	\end{axis}
	\end{tikzpicture}
	\caption{Microarchitecture Utilization.}\label{swcpi}
\end{figure}

Surprisingly, the AVX512 version of the Scan algorithm shows no improvement over the AVX2 implementation. To explain this we need to take a closer look at the microarchitecture of the vector pipeline. Every vector core in Intel processors have 7 ports tuned for the execution of the certain class of instructions \cite{b6}. In the absence of data dependencies between the instructions, the core is capable of executing up to 4 commands per the CPU cycle so the minimum (best) achievable cycle-per-instruction ratio is 0.25. The CPI rate is a good metric of the instruction level parallelism, so we have recorded the CPI measurements for our experimental run, as presented in Figure \ref{swcpi}. 

The bulk of SW computation consists of additions and comparisons (max operations).  These instructions could execute on ports 0, 1, and 5. Additionally, port 5 handles all vector shift operations of the striped algorithm. So it is convenient to model the SW instruction parallelism simply as the distribution of the arithmetic instructions across ports 0 and 1 with CPI approximately 0.5. Potential increase in CPI due to stall caused by the data dependencies is roughly compensated by the additional shuffle instructions executed on port 5. Numbers for the SSE and AVX2 implementations support this simplistic model.

The CPU in our experimental setup is build on the Skylake microarchitecture. In Skylake, port 5 is the only fully built port capable of processing 512-bit wide vectors. Ports 0 and 1 are still 256 bit wide, and for handling the 512 bit vectors they are fused together to form a single port 0+1. So in our SW execution model, transition to AVX512 simply trades executing the arithmetic instructions across two ports 0 and 1 for a single double wide port 0+1 with CPI rising to 1 as clear from the graph. Obviously, the overall throughput does not change and neither does the run time.

After such a disappointing result we have redoubled our efforts to produce better solution on the AVX512 microarchitecture. Turns out that using only 256-bit wide instructions from the AVX512VL set, along with masking and utilizing all 32 available vector registers, keeps ports 0 and 1 separate and yields the best execution times.  CPI is still quite high due the masked instructions executing only on port 5 and clogging it further.

The last observation is that CPI of the scan algorithm is consistently higher that CPI of the lazy-F. Implementing vector scan requires a long sequence of shuffle and arithmetic instructions depending on the completion of the previous operation. The instruction level parallelism is underutilized, leaving the scan algorithm just marginally faster even though the total number of instructions executed is significantly lower. Interleaving two SW executions on the same core could improve the data parallelism and further improve the performance of the scan algorithm in future.

\section{Conclusion}
We have enhanced the striped version of the Smith Waterman algorithm by achieving better asymptotic and practical performance, and by future-proofing it for the increasing width vector architectures. Our version combines the best features of scan and striped approaches making it equally useful for all sequence lengths and vector width.

Modern CPUs allow for multiple levels of parallel execution, and achieving top performance requires all levels working in coordination. We have augmented the data parallel vector approach with the simple model for instruction level parallelism that has provided practical insights and directions for further advancements.

\vspace{12pt}
\vfill
\end{document}